\documentclass[final,3p,times,twocolumn]{elsarticle}
\usepackage{amsmath}
\usepackage{amsfonts}
\usepackage{amssymb}
\usepackage{slashed}
\usepackage{graphicx}
\newcounter{bla}

\journal{Computer Physics Communications}

\begin{document}
	\begin{frontmatter}	
		
		\title{Quark propagator with complex-valued momentum from Schwinger-Dyson equation in the Euclidean space}
		
		\author[a]{Shaoyang Jia\corref{author}}
		\author[a]{Ian Clo\"{e}t}
		
		\cortext[author] {Corresponding author.\\\textit{E-mail address:} syjia@anl.gov}
		\address[a]{Physics Division, Argonne National Laboratory, 9700 S. Cass Avenue, Lemont, IL 60439, USA}
		\begin{abstract}			
			In the Euclidean-space formulation of integral equations for the structure of quantum chromodynamics (QCD) bound states, the quark propagators with complex-valued momentum are densely sampled. We therefore propose an accurate and efficient algorithm to compute these propagators. The quark propagator both on the spacelike real axis and at complex-valued momenta is determined from its Schwinger-Dyson equation (SDE). We first apply an iterative solver to determine the quark propagator on the spacelike real axis. The propagator at complex-valued momenta is then computed from its self-energy based on this solution, where demanding integrals are encountered. In order to compute of these integrals, we apply customized variable transformations for the radial integral after subtracting the asymptotics. We subsequently apply an optional compound of quadrature rules for the angular integral. The contribution from the asymptotics is added at the last step. The accuracy and the performance of this algorithm for the quark propagator at complex-valued momentum are tested in comparison with an adaptive quadrature. 
		\end{abstract}
		\begin{keyword}
			fixed-grid quadrature \sep quark propagator \sep Schwinger-Dyson equation.
		\end{keyword}
	\end{frontmatter}
	%
	{\bf PROGRAM SUMMARY}\\
	\begin{small}
		\noindent
		{\em Program Title:} \texttt{SDE\_quark} \\
		{\em CPC Library link to program files:} (to be added by Technical Editor) \\
		{\em Developer's repository link:} (if available) \\
		{\em Code Ocean capsule:} (to be added by Technical Editor) \\
		{\em Licensing provisions:} \texttt{GPLv3} \\
		{\em Programming language:} \texttt{C++} \\
		{\em Nature of problem:}\\
		Integral equations for structure of QCD bound states require quark propagators with complex-value momentum as inputs. These propagators are solved from their Schwinger-Dyson equation (SDE) as nonlinear integral equations. Solutions can be obtained from an iterative solver for positive real momenta. When computing the propagators with complex momentum based on the real solutions, integrations in the SDE are difficult to evaluate numerically. \\
		{\em Solution method:}\\
		The order of integration matters when the radial integrals in the SDE contain UV divergence. The contribution to the radial integrals from the asymptotics is handled separately to facilitate convergence. Tailored variable transformations and quadrature rules are constructed to compute the radial integrals and angular integrals in the SDE for the quark propagators with complex-valued momentum. \\
		{\em Additional comments including restrictions and unusual features:}\\
		The program has been utilized in the Bethe-Salpeter equation (BSE) for mesons. Numerical tolerances, branching conditions for quadrature rules, and variable transformations can be adjusted. Application is not limited by the poles of the quark propagator nearest to the positive real axis because the the inverse of the quark propagator is evaluated. Multiprocessing can be enabled through \texttt{OpenMP}. Instructions to build optional compiled \texttt{pybind11} modules that allow the program to run in \texttt{Python} are provided~\cite{1}. 
		
	\end{small}
	%
	%
	\section{Introduction}
	In terms of Green's functions of a quantum field theory, the structure of mesons and baryons as relativistic bound states of quarks and gluons due to quantum chromodynamics (QCD) interactions are solved from the Bethe-Salpeter equations (BSEs) and the Faddeev equations~\cite{Maris:1999nt,Eichmann:2016hgl,Sanchis-Alepuz:2017jjd}. When these equations are formulated in the Euclidean space after the Wick rotation, one essential input is the propagator of the quarks with complex-valued momenta. Another input is the interaction of constituents, a well-established truncation of which is the Maris-Tandy model~\cite{Maris:1999nt}. After adopting this truncation, propagators of quarks are solved from their Schwinger-Dyson equation (SDE) in the Euclidean space on the positive real axis for spacelike momenta. Based on the solutions obtained, their values at complex momenta can subsequently be calculated~\cite{Windisch:2016iud}. When solving the integral equations for the structure of bound states numerically, frequent requests of the quark propagators at complex momenta are made. Therefore it is crucial to have an accurate and efficient method for their computation. 
	
	After the Wick rotation, the SDE for the quark propagator becomes a set of integral equations for real-valued scalar functions of positive real momenta. Integrals in this SDE are straightforward to evaluate numerically when applying the Maris-Tandy model. We therefore provide an iterative solver for the resulting integral equations. The same integrals are however difficult to compute numerically when the quark momentum is complex.
	
	Although one could evaluate quark propagators along a boundary of their holomorphic regions in the complex-momentum plane to apply the Cauchy integration formula numerically for their values enclosed, such a method depends on the accuracy of numerical loop integrals in the complex-momentum plane~\cite{Dorkin:2013rsa}. Aside from requesting the propagators at complex momenta in the first place, this method cannot go beyond singularities nearest to the positive real axis. We propose an algorithm that applies tailored variable transformations and quadrature rules to compute quark propagators at complex momenta from their SDE instead. Doing so also avoids the difficulty of sampling singularities in the propagators. 
	
	This article is organized as follows. Section~\ref{sc:sde_euc} discusses the SDE for the quark propagator in the Euclidean space, its iterative solver, and the asymptotic contribution to the radial integrals. Section~\ref{sc:quad_and_vt} includes the implementation of variable transformations and quadrature rules to evaluate the quark propagator with complex momentum. Testing results of the program are given in Section~\ref{sc:test}. Quadrature rules are described in the Appendices. 
	\section{SDE for the quark propagator in the Euclidean space with rainbow-ladder truncation\label{sc:sde_euc}}
	\subsection{SDE as nonlinear integral equations}
	\begin{figure}
		\centering
		\includegraphics[width=\linewidth]{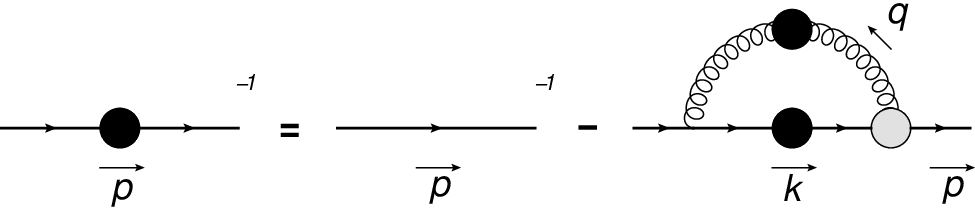}
		\caption{SDE for the quark propagator. Solid lines are the quark propagators. While the curved line is the gluon propagator. Black blobs indicate dressed propagators. The gray blob is the quark-gluon vertex. In the Maris-Tandy model, this vertex combined with the dressed gluon propagator are replaced by the bare vertex and an Ansatz gluon propagator in the Landau gauge.}
		\label{fig:sdequark}
	\end{figure}
	We work in the $4$-dimensional Euclidean space where $p^2\geq 0$ for spacelike $p_j$ with ${j\in\{1,2,3,4\}}$. The dressed quark propagator decomposes into its Dirac vector and Dirac scalar components according to
	\begin{equation}
		S_{\mathrm{F}}(p) = - \slashed{p}\,\sigma_{\mathrm{v}}(p^2) + \sigma_{\mathrm{s}}(p^2),\label{eq:def_sigma_vs}
	\end{equation}
	which is determined from its SDE diagrammatically represented in Fig.~\ref{fig:sdequark}. We have defined ${\slashed{p} = \sum_{j=1}^{4}\gamma_j p_j}$ with $\gamma_j$ being the Dirac matrices in the Euclidean space.	Within the Maris-Tandy model, the dressed gluon propagator is combined with the quark-gluon vertex such that
	\begin{equation}
		\sum_{i=1}^{4}\Gamma_i(k, p)\, D_{ij}(q^2) = ( \gamma_j - \slashed{q} q_j/q^2 )\, \mathcal{G}(q^2)
	\end{equation}
	in the Landau gauge~\cite{Maris:1999nt}. Here ${\mathcal{G}(q^2)}$ is the dressing function for the gluon propagator. This function can be any alternative gluon dressing function in the rainbow-ladder truncation, where the Dirac structure of the quark-gluon vertex is given by $\gamma_j$. We can similarly decompose the inverse of the quark propagator of momentum $p_j$ into Dirac components:
	\begin{equation}
		S_{\mathrm{F}}^{-1}(p) = \slashed{p}\,A(p^2) + B(p^2).\label{eq:def_A_B}
	\end{equation}
	These components with real and spacelike $p^2$ are solutions of the SDE we would like to find in the Euclidean space. Let us first define the denominator as
	\begin{equation}
		\mathcal{D}(p^2) = p^2 \, A^2(p^2) + B^2(p^2).
	\end{equation}
	After representing $p^2$ by either $x$ or $y$ for notational convenience, based on Eqs.~(15) and (16) of Ref.~\cite{Windisch:2016iud} we obtain
	\begin{subequations}\label{eq:sde_quark_prop}
		\begin{align}
			A(x) & = Z_2 + \dfrac{1}{6\pi^3} \int_{-1}^{1}dz\,\sqrt{ 1 - z^2 } \, \int_{0}^{\Lambda^2_{\mathrm{UV}}}dy\, \dfrac{ y\, A(y)}{ \mathcal{D}( y ) } \nonumber\\
			& \quad \times K(x,y,z) \, \mathcal{G}( x + y - 2z \sqrt{xy}), \\
			B(x) & = Z_2 m_{\mathrm{B}} + \dfrac{1}{6\pi^3} \int_{-1}^{1}dz\,\sqrt{ 1 - z^2 } \, \int_{0}^{\Lambda^2_{\mathrm{UV}}}dy\, \dfrac{ 3\,y\, B(y)}{ \mathcal{D}( y ) } \nonumber\\
			& \quad \times \mathcal{G}( x + y - 2z \sqrt{xy}),
		\end{align}
	where the kernel function is defined as
		\begin{equation}
			K(x,y,z) = 2z\sqrt{y/x} + \dfrac{ \left(1 + y/ x\right)z\sqrt{xy} - 2 y }{ x + y - 2z\sqrt{xy}}.
		\end{equation}
	\end{subequations}
	With $4$-dimensional spacetime, renormalization is required for Eq.~\eqref{eq:sde_quark_prop}. The bare mass $m_{\mathrm{B}}$ is related to the renormalized mass $m_{\mathrm{R}}$ through $m_{\mathrm{B}} = Z_m m_{\mathrm{R}}$. The renormalization constants $Z_2$ and $Z_{\mathrm{m}}$ are determined from renormalization conditions. We choose these conditions such that the quark propagator is given by its free-particle form with the renormalized mass at the renormalization point $\mu^2$. Explicitly we have
	\begin{equation}
		\begin{cases}
			A(\mu^2) = 1.0\\
			B(\mu^2) = m_{\mathrm{R}} 
		\end{cases}. \label{eq:rnm_cond}
	\end{equation}
	In terms of the inverse propagator in Eq.~\eqref{eq:def_A_B}, the Dirac components of the propagator in Eq.~\eqref{eq:def_sigma_vs} are given by
	\begin{equation}
		\begin{cases}
			\sigma_{\mathrm{v}}(p^2) = A(p^2) / \mathcal{D}(p^2) \\
			\sigma_{\mathrm{s}}(p^2) = B(p^2) / \mathcal{D}(p^2)
		\end{cases},\label{eq:sigma_vs_from_AB}
	\end{equation}
	which can be demonstrated after multiplying Eq.~\eqref{eq:def_sigma_vs} with Eq.~\eqref{eq:def_A_B} to get unity. Notice that in the Euclidean space we have $\slashed{p}\slashed{p} = -p^2$ instead of $p^2$ in the Minkowski space. 
	
	There remain the radial integrals for the variable $y$ and the angular integrals for the variable $z$ of the loop momentum in Eq.~\eqref{eq:sde_quark_prop}. If one applies the tree-level Feynman rules with the perturbative gluon propagator, the quark self-energy is logarithmically divergent, which is not the case from a naive power counting of the radial integrals in Eq.~\eqref{eq:sde_quark_prop}. Consistency with the perturbative analysis can be restored by symmetrize the kernels of the integrals with respect to $z = 0$. We therefore develop our algorithm to compute the quark propagator based on the following equivalent form of Eq.~\eqref{eq:sde_quark_prop}:
	\begin{subequations}\label{eq:sde_quark_prop_symmeetric_z}
		\begin{align}
			& A(x) = Z_2 + \dfrac{1}{3\pi^3} \int_{0}^{1}dz\,\sqrt{ 1 - z^2 } \, \int_{0}^{\Lambda^2_{\mathrm{UV}}}dy\, \dfrac{ y\, A(y)}{ 4 \mathcal{D}( y ) } \nonumber\\
			& \quad \times \Big\{ \left[ K(x,y,z) + K(x,y,-z) \right] \nonumber\\ 
			& \quad \times \left[ \mathcal{G}( x + y - 2z \sqrt{xy}) + \mathcal{G}( x + y + 2z \sqrt{xy}) \right] \nonumber\\
			& \quad + \left[ K(x,y,z) - K(x,y,-z) \right] \nonumber\\
			& \quad \times \left[ \mathcal{G}( x + y - 2z \sqrt{xy}) - \mathcal{G}( x + y + 2z \sqrt{xy}) \right] \Big\},\\
			& B(x) = Z_2 m_{\mathrm{B}} + \dfrac{1}{3\pi^3} \int_{0}^{1}dz\,\sqrt{ 1 - z^2 } \, \int_{0}^{\Lambda^2_{\mathrm{UV}}}dy\, \dfrac{ 3\,y\, B(y)}{ 2 \mathcal{D}( y ) } \, \nonumber\\
			& \quad\times \left[ \mathcal{G}( x + y - 2z \sqrt{xy}) + \mathcal{G}( x + y + 2z \sqrt{xy} ) \right].
		\end{align}
	\end{subequations}
	Nonperturbatively the radial integrals in Eq.~\eqref{eq:sde_quark_prop_symmeetric_z} are only convergent when the gluon propagator $\mathcal{G}(q^2)$ decays faster than $\mathcal{O}(1/q^2)$ asymptotically. Otherwise as in the case of perturbative QCD calculations, the radial integrals are ultraviolet (UV) divergent, mandating either a momentum cutoff $\Lambda^2_{\mathrm{UV}}$ or any other regularization schemes to be applied to obtain finite radial integrals. 
	
	With real and positive $x$, Eq.~\eqref{eq:sde_quark_prop_symmeetric_z} forms a set of nonlinear integral equations for the real-valued scalar functions $A(x)$ and $B(x)$. Once a gluon propagator and a set of renormalization conditions are given, these equations can be solved using an iterative algorithm starting from an initial guess. For the iterative solver, it is convenient to compute the angular integrals first as it involves only known functions. Because the right-hand sides of Eq.~\eqref{eq:sde_quark_prop_symmeetric_z} only sample the inverse of the quark propagator with positive real momentum $y$, solutions from the iterative solver can subsequently be used to determine the value of the quark propagator in the complex-momentum plane of $x$. 
	
	When divergence is present for the radial integrals of Eq.~\eqref{eq:sde_quark_prop_symmeetric_z} with large UV cutoffs, the order of integrals taken in Eq.~\eqref{eq:sde_quark_prop_symmeetric_z} matters as the series of functions of $z$ representing any consecutive numerical approximations of the radial integral is not uniformly convergent when $x$ is complex valued. An optimized algorithm to compute integrals in Eq.~\eqref{eq:sde_quark_prop_symmeetric_z} for complex-valued $x$ should handle the radial integrals before the angular ones. One exception exists when $x$ is real and positive, which corresponds to the iterative solver of the SDE in the Euclidean space. We also provide the option of Pauli-Villars regularization in addition to the UV cutoff~\cite{Maris:1999nt}. 
	\subsection{Iterative solver in the Euclidean space}
	Having converted the SDE for the quark propagator in the rainbow-ladder truncation into integral equations in Eq.~\eqref{eq:sde_quark_prop_symmeetric_z}, its numerical solution requires a gluon propagator as an input. We adopt the well-established Maris-Tandy model~\cite{Maris:1999nt}, where the gluon propagator is written as a sum of an inferred (IR) term and a UV term:
	\begin{subequations}\label{eq:def_MT_model}
		\begin{align}
			& \mathcal{G}(q^2) = \mathcal{G}_{\mathrm{IR}}(q^2) + [ 1 - \exp ( - q^2 / (4m^2_{\mathrm{t}}) ) ] / q^2 \nonumber\\
			& \quad \times \dfrac{2\,c_{\mathrm{g}}}{ \ln\left[ e^2-1+\left(1 + q^2/\Lambda_{\mathrm{QCD}}^2 \right)^2 \right]},
		\end{align}
	with 
		\begin{equation}
			\mathcal{G}_{\mathrm{IR}}(q^2) = \dfrac{4\pi D}{\omega^6}q^2\, \exp( -q^2/\omega^2 ) \label{eq:def_mt_IR}
		\end{equation}
	\end{subequations}
	and ${c_{\mathrm{g}} = 48\pi^2/(33-2N_{\mathrm{f}})}$. For pions, kaons, and rho mesons, a typical choice of model parameters is given in Table~\ref{tab:pmts_MT}. 
	
	We would like to solve Eq.~\eqref{eq:sde_quark_prop_symmeetric_z} for spacelike $x$. The maximum reach of the quark momentum is $\Lambda^2_{\mathrm{UV}}$ applying the cutoff regularization. Therefore a numerical grid for ${x\in[0,\Lambda^2_{\mathrm{UV}}]}$ needs to be constructed. Specifically we first map the variable $x$ by
	\begin{equation}
		u(x) = x / ( x + C ),\label{eq:def_u_of_p2}
	\end{equation}
	with $C$ being the scale of the transformation. This variable transform results in an inverse transform of ${x = C u / ( 1 - u )}$ and the following change of integration measure:
	\begin{equation}
		\int_{0}^{\Lambda^2_{\mathrm{UV}}}dx = \int_{0}^{u_{\mathrm{max}}}du\,C/(1-u)^2. 
	\end{equation}
	The variable $u$ falls within $[0,u_{\mathrm{max}}]$ with ${u_{\mathrm{max}}} = {u(\Lambda^2_{\mathrm{UV}})}$. We subsequently discretize the variable $u$ in an $n$-point Gauss-Chebyshev quadrature grid of the first kind with the assistance of a linear map to $[-1,1]$, which is described in \ref{sc:quad_gc}. This quadrature rule is chosen due to the expected divergence when the upper end of the integration corresponds to infinite $p^2$. 
	
	After choosing the grid for the mapped momentum variable, we would like to compute the kernel of the SDE for an iterative solver. Specifically for each given pair of positive $x$ and $y$, we apply the adaptive Gauss-Kronrod quadrature described in \ref{sc:quad_gk} to evaluate the integral with respect to the angular variable $z$. The kernel of integrations in Eq.~\eqref{eq:sde_quark_prop_symmeetric_z} then becomes a $2$-dimensional array of size $n$ for each scalar component. In order to compute the self-energy at the renormalization scale $\mu^2$, there is an additional vector of size $n$ for each of the scalar components. These kernel arrays are independent of the Euclidean-space solution therefore generated and stored in the memory. 
	
	The initial guess for the inverse of the quark propagator is
	${A(x) = 1}$ and ${B(x) = m_{\mathrm{R}}}$, with $m_{\mathrm{R}}$ being the input renormalized mass. At each step in iteration, the inverse of the quark propagator from the previous iteration is used 
	to obtain the new self-energy based on Eq.~\eqref{eq:sde_quark_prop_symmeetric_z}. The renormalization constants are subsequently updated based on Eqs.~\eqref{eq:rnm_cond} and~\eqref{eq:sde_quark_prop_symmeetric_z}. Relative and absolute changes in both the renormalization constants and the inverse propagator are tested for the desired tolerances. The iterative solver stops if the tolerance conditions are satisfied. Otherwise another step of iteration is taken. Once the numerical solution has been obtained, the inverse of the quark propagator with positive real momenta below the UV cutoff and not at the solution grid points is computed from the natural cubic spline interpolation~\cite{galassi2009gnu}. The inverse propagator above the UV cutoff is approximated by its values at the cutoff. 
	\subsection{Asymptotic behavior of the radial integrals}
	\begin{table}
		\centering
		\begin{tabular}{ccccc}
			\hline
			$\omega$ & $D$ & $N_{\mathrm{f}}$ & $\Lambda_{\mathrm{QCD}}$ & $m_{\mathrm{t}}$ \\
			\hline
			$0.4~\mathrm{GeV}$ & $0.859~\mathrm{GeV}^2$ & $4$ & $0.234~\mathrm{GeV}$ & $0.5~\mathrm{GeV}$ \\
			\hline
		\end{tabular}
		\caption{A typical choice of model parameters in the Maris-Tandy model interaction for light mesons.}\label{tab:pmts_MT}
	\end{table}
	The gluon propagator of the Maris-Tandy model in Eq.~\eqref{eq:def_MT_model} is dominated by the UV term asymptotically. For the computation of the quark propagator with complex-valued momentum, we would like to isolate the contribution to the quark self-energy from asymptotic terms. Despite the implementation of the UV cutoff for the radial integrals, computing the  asymptotic contribution separately from the UV-finite part facilitates convergence. Specifically we obtain the following limit
	\begin{equation}
		\lim\limits_{q^2\rightarrow +\infty} \mathcal{G}(q^2) = \dfrac{ c_{\mathrm{g}} }{ q^2 \ln \left(q^2/\Lambda_{\mathrm{QCD}}^2 \right) }. \label{eq:asymp_MT_model}
	\end{equation}
	Notice that the kernel functions in Eq.~\eqref{eq:sde_quark_prop_symmeetric_z} tend towards constants for large spacial momentum $y$. The asymptotic contribution to the radial integrals in Eq.~\eqref{eq:sde_quark_prop_symmeetric_z} is explicitly given by
	\begin{subequations} \label{eq:div_quark_self_energy}
		\begin{align}
			& \dfrac{ 2c_{\mathrm{g}} }{ A(\Lambda^2_{\mathrm{UV}}) }\int_{0}^{1}dz \sqrt{1-z^2} \int_{}^{\Lambda_{\mathrm{UV}}^2}dy\, \big[ -2 ( 1- z^2) \nonumber\\
			& \quad + 6z^2 \left(1 + 1 / \ln y \right) \big] \dfrac{1}{y\ln y}, \\
			& \dfrac{ 2c_{\mathrm{g}} B(\Lambda^2_{\mathrm{UV}})}{ A^2(\Lambda^2_{\mathrm{UV}}) } \int_{0}^{1}dz\sqrt{1-z^2} \int_{}^{\Lambda_{\mathrm{UV}}^2}dy\, \dfrac{3}{y\ln y },
		\end{align}
	\end{subequations}
	with the lower limits of the integration anywhere between $0$ and $\Lambda^2_{\mathrm{UV}}$. The limits of $A(y)$ and $B(y)$ at large spacelike $y$ are approximated by their values at the UV cutoff. Integrals in Eq.~\eqref{eq:div_quark_self_energy} are independent of the external momentum $x$. They also factorize into radial and angular parts, resulting in $3$ constants to be computed in order to compensate for the subtracted asymptotic contributions. 
	
	With the asymptotics of the radial integrals in Eq.~\eqref{eq:sde_quark_prop_symmeetric_z} isolated, the remaining radial integrals can be computed through a fixed quadrature grid of a transformed variable. Explicitly based on Eq.~\eqref{eq:div_quark_self_energy}, we subtract the following functions from the integrals in Eq.~\eqref{eq:sde_quark_prop_symmeetric_z}:
	\begin{subequations} \label{eq:subtraction_quark_self_energy}
		\begin{align}
			& C_{\mathrm{v}}(y, z, y_0) = \dfrac{c_{\mathrm{g}}}{ A(\Lambda^2_{\mathrm{UV}}) } \, \left( \dfrac{y}{y+y_0} \right)^2 \Bigg[ -2 ( 1- z^2 ) + 6z^2 \nonumber\\
			& \quad \times \left(1 + \dfrac{1}{\ln(e + y/y_0)} \right) \Bigg] \dfrac{1}{(y+y_0) \ln (e + y/y_0)}, \\
			& C_{\mathrm{s}}(y, z, y_0) = \dfrac{c_{\mathrm{g}} B(\Lambda^2_{\mathrm{UV}})}{ A^2(\Lambda^2_{\mathrm{UV}}) } \dfrac{3y}{(y+y_0)^2\ln(e + y/y_0) },
		\end{align}
	\end{subequations}
	with ${y_0 = \Lambda^2_{\mathrm{QCD}}}$. The term $y_0$ is introduced to make the subtracted kernel well-defined near ${y=0}$, avoiding complexities in the small $y$ region. These subtractions require the following constants to be added after integrations with respect to $z$ and $y$ in order to restore the original results in Eq.~\eqref{eq:sde_quark_prop_symmeetric_z}:
	\begin{subequations} \label{eq:div_quark_self_energy_profile}
		\begin{align}
			& c_{\mathrm{v}}(y_0, \Lambda^2_{\mathrm{UV}}) = \dfrac{ 2c_{\mathrm{g}} }{ A(\Lambda^2_{\mathrm{UV}}) }\int_{0}^{1}dz \sqrt{1-z^2} \int_{0}^{\Lambda_{\mathrm{UV}}^2}dy  \dfrac{y^2}{(y+y_0)^3} \nonumber\\
			& \quad \times \left[ -2 ( 1- z^2) + 6z^2 \left(1 + \dfrac{1}{\ln(e + y/y_0)} \right) \right] \dfrac{1}{\ln (e + y/y_0)} \nonumber\\
			& = \dfrac{3\pi \,c_{\mathrm{g}}}{4 A(\Lambda^2_{\mathrm{UV}})} \left[ - t_{1} (y_0, \Lambda^2_{\mathrm{UV}}) + t_{2}(y_0, \Lambda^2_{\mathrm{UV}}) \right], \\
			& c_{\mathrm{s}}(y_0, \Lambda^2_{\mathrm{UV}}) = \dfrac{ 2c_{\mathrm{g}}B(\Lambda^2_{\mathrm{UV}})}{ A^2(\Lambda^2_{\mathrm{UV}}) } \int_{0}^{1}dz\sqrt{1-z^2} \int_{0}^{\Lambda_{\mathrm{UV}}^2}dy \nonumber\\
			& \quad \times \dfrac{3y}{(y+y_0)^2\ln(e + y/y_0) }	= \dfrac{ 3\pi\, c_{\mathrm{g}} B(\Lambda^2_{\mathrm{UV}})}{ 2 A^2(\Lambda^2_{\mathrm{UV}}) } t_{0}(y_0,\Lambda^2_{\mathrm{UV}}).
		\end{align}
	Here we have defined the following constants
		\begin{align}
			t_{0}(y_0,\Lambda^2_{\mathrm{UV}}) & = \int_{0}^{\Lambda_{\mathrm{UV}}^2}dy\, \dfrac{y}{(y+y_0)^2 \ln( e + y / y_0)}, \\
			t_{1}(y_0,\Lambda^2_{\mathrm{UV}}) & = \int_{0}^{\Lambda_{\mathrm{UV}}^2}dy\, \dfrac{y^2}{(y+y_0)^3 \ln( e + y / y_0)}, \\
			t_{2}(y_0,\Lambda^2_{\mathrm{UV}}) & = \int_{0}^{\Lambda_{\mathrm{UV}}^2}dy\, \dfrac{y^2 \left[ 1 + 1/\ln ( e + y / y_0 ) \right] }{(y+y_0)^3 \ln( e + y / y_0)} .
		\end{align}
	\end{subequations}
	We apply the adaptive Gauss-Kronrod quadrature for their computation directly on the integration variable $y$. The expressions for the SDE of the quark propagator in Eq.~\eqref{eq:sde_quark_prop_symmeetric_z} then become
	\begin{subequations}\label{eq:sde_quark_prop_final_numeric}
		\begin{align}
			& A(x) = Z_2 + \dfrac{1}{6\pi^3} \Bigg\{c_{\mathrm{v}}(y_0,\Lambda^2_{\mathrm{UV}}) +  2\int_{0}^{1}dz \sqrt{ 1 - z^2 } \nonumber\\
			& \times \int_{0}^{\Lambda^2_{\mathrm{UV}}}dy \,\bigg\{ \dfrac{ y\, A(y)}{ 4 \mathcal{D}( y ) } \Big\{ \left[ K(x,y,z) + K(x,y,-z) \right] \nonumber\\ 
			& \times \left[ \mathcal{G}( x + y - 2z \sqrt{xy}) + \mathcal{G}( x + y + 2z \sqrt{xy}) \right] \nonumber\\
			& + \left[ K(x,y,z) - K(x,y,-z) \right]\, \big[ \mathcal{G}( x + y - 2z \sqrt{xy}) \nonumber\\
			& - \mathcal{G}( x + y + 2z \sqrt{xy}) \big] \Big\} - C_{\mathrm{v}}(y,z,y_0) \bigg\} \Bigg\}, \label{eq:sde_quark_prop_final_numeric_vec}
		\end{align}
	and 
		\begin{align}
			& B(x) = Z_2 m_{\mathrm{B}} + \dfrac{1}{6\pi^3} \Bigg\{ c_{\mathrm{s}}(y_0,\Lambda^2_{\mathrm{UV}}) + 2\int_{0}^{1}dz \sqrt{ 1 - z^2 } \nonumber\\
			& \times \int_{0}^{\Lambda^2_{\mathrm{UV}}}dy\, \bigg\{ \dfrac{ 3\,y\, B(y)}{ 2\mathcal{D}( y ) } \big[ \mathcal{G}( x + y - 2z \sqrt{xy}) \nonumber\\
			& + \mathcal{G}( x + y + 2z \sqrt{xy}) \big] - C_{\mathrm{s}}(y,z,y_0) \bigg\} \Bigg\}, \label{eq:sde_quark_prop_final_numeric_sec}
		\end{align}
	\end{subequations}
	after separating contributions from the asymptotics. 
	\section{Fixed-grid quadrature for the quark self-energy\label{sc:quad_and_vt}}
	\subsection{Radial integrals}
	Having obtained the renormalization constants and the quark propagator with positive real momenta from the iterative solver of the SDE, we would like to develop a scheme to compute the propagator at complex-valued momenta based on fixed-grid quadrature rules. These rules are to be applied to determine the inverse of the quark propagator through Eq.~\eqref{eq:sde_quark_prop_final_numeric}, where contributions from the asymptotics are computed separately from the rest of the gluon propagator. The propagator is then obtained from its inverse utilizing Eq.~\eqref{eq:sigma_vs_from_AB}. 
	\paragraph{Separation of contribution from the IR term}
	With ${x\in\mathbf{C}}$ being the momentum of the quark propagator, in the limit of ${\Re\{x\} \gg 1.0\,\mathrm{GeV}^2}$ and 
	${z\,\Re\{\sqrt{x}\} \geq \omega}$ the IR term in Eq.~\eqref{eq:def_MT_model} becomes extremely close to a $\delta$-function. In this scenario, contributions from the IR term can be easily missed numerically by a quadrature rule designed to account for features of the UV term. Also due to the drastically different asymptotic behaviors of the UV and the IR terms, we compute their contributions to the quark self-energy separately in this scenario.
	
	In order to center and magnify the exponential profile of the IR term, we introduce the following variable transformation:
	\begin{equation}
		u = \left( \sqrt{y}- z\,\Re\{\sqrt{x}\} \right) / \omega, \label{eq:def_vt_u}
	\end{equation}
	the inverse of which is	given by ${y = (\omega u + z\,\Re\{\sqrt{x}\})^2}$. The contribution to radial integrals from the IR term can then be evaluated applying the Gauss-Hermite quadrature rule considering the presence of the exponential factor. Because the Gauss-Hermite quadrature samples to infinity, the finite UV cutoff of the integration is accounted by the following identity
	\begin{equation}
		\int_{0}^{\Lambda^2_{\mathrm{UV}}}dy\,f(y) = \int_{0}^{+\infty}dy\, \left[ f(y) - f(y+\Lambda^2_{\mathrm{UV}}) \right]. 
	\end{equation}
	The application of Eq.~\eqref{eq:def_vt_u} converts the integral with respect to $y$ according to
	\begin{align}
		& \int_{0}^{\Lambda^2_{\mathrm{UV}}}dy\, f(y) = \int_{-z\,\Re\{\sqrt{x}\}/\omega}^{+\infty}du\, 2\omega\, ( \omega u + z\,\Re\{\sqrt{x}\} ) \nonumber\\
		& \times \left[ f\left( ( \omega u + z\,\Re\{\sqrt{x}\} )^2 \right) - f\left( ( \omega u + z\,\Re\{\sqrt{x}\} )^2 +\Lambda^2_{\mathrm{UV}} \right)\right] \nonumber\\
		& = \int_{-\infty}^{+\infty}du\, 2\omega \, ( \omega u + z\,\Re\{\sqrt{x}\}) \, \theta\left( \omega u + z\,\Re\{\sqrt{x}\} \right) \nonumber\\
		& \times \left[ f\left( ( \omega u + z\,\Re\{\sqrt{x}\} )^2 \right) - f\left( ( \omega u + z\,\Re\{\sqrt{x}\} )^2 +\Lambda^2_{\mathrm{UV}} \right) \right] \nonumber\\
		& \simeq \sum_{j=1}^{N_{\mathrm{GH}}} 2w_{j}\, \omega\, ( \omega u_{j} + z\,\Re\{\sqrt{x}\}) \, \theta\left( \omega u_{j} + z\,\Re\{\sqrt{x}\} \right) \nonumber\\
		& \times \Big[ f\left( ( \omega u_{j} + z\,\Re\{\sqrt{x}\} )^2 \right) \nonumber\\
		& \quad - f\left( ( \omega u_j + z\,\Re\{\sqrt{x}\} )^2 +\Lambda^2_{\mathrm{UV}} \right) \Big], \label{eq:def_gauss_hermite}
	\end{align}
	with $u_j$ and $w_j$ being the roots and weights of the Gauss-Hermite quadrature rule of $N_{\mathrm{GH}}$ points as specified in \ref{sc:quad_gh}. Notice that exponential factors of $\exp( u^2_j )$ are absorbed into the weights $w_j$. 
	
	The IR contribution to the radial integrals is computed separately when the following conditions are satisfied
	\begin{equation}
		\begin{cases}
			\Re\{x\} \geq r_{\mathrm{threshold}} = 7.5~\,\mathrm{GeV}^2 \\[1mm]
			z\,\Re\{\sqrt{x}\} / \omega \geq h_{\mathrm{threshold}} = 2.0
		\end{cases}. \label{eq:conditions_gauss_hermite}
	\end{equation}
	We find that under these conditions the Gaussian profile of the IR term in Eq.~\eqref{eq:def_MT_model} becomes narrow with respect to the other scales of the integral, prompting the use of the Gauss-Hermite quadrature to compute its contribution. Specifically Eq.~\eqref{eq:def_gauss_hermite} is applied to evaluate the radial IR contribution to Eq.~\eqref{eq:sde_quark_prop_final_numeric} with 
	\begin{subequations}
		\begin{align}
			& f_{A}(y) = \dfrac{ y\, A(y)}{ 4\mathcal{D}( y ) }
			\Big\{ [ K(x,y,z) + K(x,y,-z) ] \nonumber\\
			& \times [ \mathcal{G}_{\mathrm{IR}}( x + y - 2z \sqrt{xy}) + \mathcal{G}_{\mathrm{IR}}( x + y + 2z \sqrt{xy}) ] \nonumber\\
			& + [ K(x,y,z) - K(x,y,-z) ] \, [ \mathcal{G}_{\mathrm{IR}}( x + y - 2z \sqrt{xy}) \nonumber\\
			& \quad - \mathcal{G}_{\mathrm{IR}}( x + y + 2z \sqrt{xy}) ] \Big\},
		\end{align}
	and
		\begin{align}
			& f_{B}(y) = \dfrac{ 3\,y\, B(y)}{ 2 \mathcal{D}( y ) } [ \mathcal{G}_{\mathrm{IR}}( x + y - 2z \sqrt{xy}) \nonumber\\
			& + \mathcal{G}_{\mathrm{IR}}( x + y + 2z \sqrt{xy}) ]
		\end{align}
	\end{subequations}
	respectively for the Dirac vector and scalar components in the inverse of the self-energy. Because the asymptotics is dominated by the UV term, the subtraction in Eq.~\eqref{eq:sde_quark_prop_final_numeric} are not applied. After computing the IR contribution, the remaining radial integral can be evaluated using the following method. 
	\paragraph{Variable transform for the radial integral}
	In order to construct a fix-grid quadrature rule for the radial integrals in Eq.~\eqref{eq:sde_quark_prop_final_numeric}, let us consider the following variable transformation for a given $x\in\mathbf{C}$:
	\begin{equation}
		\theta(y) = \arccos\left( (y-c) / (y+c) \right).
	\end{equation}
	The inverse transformation is $y = c \, ( 1 + \cos \theta ) / ( 1 - \cos \theta )$, which results in a weight function of
	\begin{equation}
		w(\theta) = - 2c\,\sin(\theta) / [1 - \cos(\theta)]^2
	\end{equation}
	for the integration measure. The empirical choice for the scale of this variable transformation with the Maris-Tandy model using parameters in Table~\ref{tab:pmts_MT} is $c=1.0~\mathrm{GeV}^2$ for both components in Eq.~\eqref{eq:sde_quark_prop_final_numeric}. Applying this variable transformation for $y\in[0,\Lambda^2_{\mathrm{UV}}]$ gives
	\begin{equation}
		\int_{0}^{\Lambda^2_{\mathrm{UV}}} dy\, f(y) = \int_{\pi}^{\theta_{\mathrm{min}}} d\theta\, w(\theta)\, f\left( \dfrac{ (1 + \cos \theta)\,c }{ 1 - \cos \theta } \right) \label{eq:theta_range}
	\end{equation}
	with $\theta_{\mathrm{min}} = \theta(\Lambda^2_{\mathrm{UV}})$. 
	We then choose the Gauss-Chebyshev quadrature of the second kind with the $N_{\mathrm{GC}} = 201$ points to evaluate integrals with respect to the mapped radial variable $\theta$. This method of quadrature is chosen because the most singular contribution to the radial integral from the UV term has been handled separately in Eq.~\eqref{eq:sde_quark_prop_final_numeric}.
	
\begin{figure}
		\centering
		\includegraphics[width=\linewidth]{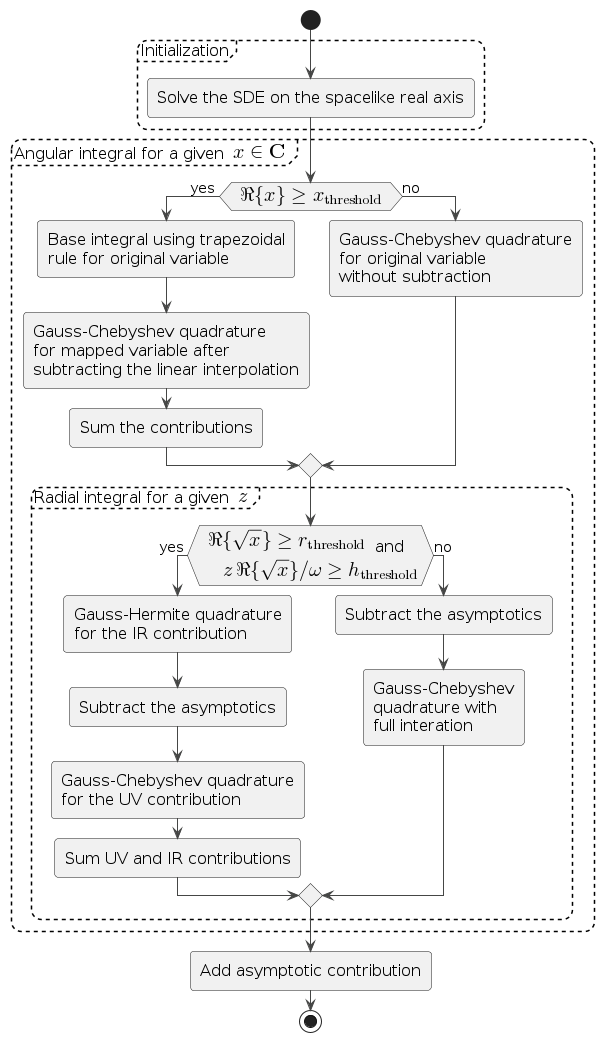}
		\caption{Activity diagram for the evaluation of the quark propagator with complex-value momentum using fixed-grid quadrature rules.}
		\label{fig:fixedgridsf}
	\end{figure}

	As a summary of computing radial integrals with asymptotics subtracted, when the condition specified in Eq.~\eqref{eq:conditions_gauss_hermite} is satisfied, we apply the Gauss-Hermite quadrature in the form of Eq.~\eqref{eq:def_gauss_hermite} to compute the IR contribution. In the meantime Eq.~\eqref{eq:theta_range} is used to compute the contribution from the UV term with Gauss-Chebyshev quadrature on the mapped variable. Otherwise when Eq.~\eqref{eq:conditions_gauss_hermite} is not satisfied, Eq.~\eqref{eq:theta_range} handles radial integrals from the full gluon propagator. This is diagrammatically represented in the radial integral section of Fig.~\ref{fig:fixedgridsf}. 
	
	It is imperative that the Gauss-Hermite quadrature is applied independently from the subsequent Gauss-Chebyshev quadrature rule for the radial integral from the UV term. In contrast to the compound of quadrature rules to be introduced for the angular integral, residues from the radial quadrature of the UV term need not be carried into the integrands of the Gauss-Hermite rule, as slight mismatches in the asymptotics may lead to large numerical error. 
	\subsection{Compound quadrature for the angular integral}
	The algorithm to compute radial integrals described in the previous subsection applies for a given value of the angular variable $z$. There remains the integral with respect to angular variable $z\in[0,1]$ in Eq.~\eqref{eq:sde_quark_prop_final_numeric} for the evaluation of the quark propagator withe complex momentum. The integrand for this angular integral consists of a base contribution augmented by a peak, the center and width of which both depend on the complex momentum $x$. Particularly when $\Re\{x\}> 0$, this peak resides narrowly and close to $z=1$ for large values of $\vert x \vert$. Despite of its presence, the $z$-integrand still vanishes at $z=1$, rendering the angular integral difficult to compute via a single fixed-grid quadrature rule. 
	
	We therefore propose the application of two quadrature rules consecutively. The first rule computes the base contribution, while the second rule handles the contribution from the peak. The base quadrature is therefore allowed to be blind to the peak of the integrand. But the interpolation upon which this quadrature rule is built needs to be simple, allowing the exact integrand to be recovered by a light-weight function. This interpolator is to be used when subtracting from the integrand a contribution that exactly matches the base quadrature. Because the subtracted integrand mainly consists of the peak in the original kernel of integration, it can be integrated with a suitable quadrature rule after a tailored variable transformation. The application of this compound quadrature depends on the condition for the real part of $x$. 
	\begin{enumerate}
		\item When $\Re\{x\} \geq x_{\mathrm{threshold}}$, we first compute the base contribution to the angular integral using the $201$-point trapezoidal rule on a uniform grid of $z\in[0,\,1]$. Both the location of the grid points and the values of the integrand sampled are passed to a linear interpolator~\cite{abramowitz1965handbook}. Such a rule is chosen because the linear interpolation of the sampled points is the exact integrand that gives a matching result for the numerical integration.
		
		Let us then define a mapped variable $\zeta$ as
		\begin{equation}
			\zeta(z) = b^{-1}\tan\left( z\,{\arctan(b)} \right),
		\end{equation}
		whose inverse map is ${ z = \arctan (b\, \zeta)\, /\, \arctan(b) }$. The weight due to this variable transformation is therefore
		\begin{equation}
			v(\zeta) = \dfrac{b}{ \arctan(b)\, ( 1 + b^2\zeta^2 ) }. 
		\end{equation}
		The limits of integration for $\zeta$ is $[0,1]$, same as those for $z$. The parameter $b$ is the root of the following transcendental equation:
		\begin{equation}
			\arctan( b \, \zeta_0 ) = z_0 \arctan( b ). \label{eq:root_finding_b}
		\end{equation}
		Here $z_0$ is roughly the center of the peak for the integrand in $z$, whose empirical expression is
		\begin{equation}
			z_0 = 1 -  d / \vert x \vert 
		\end{equation}
		for large and positive $\Re\{x\}$ with $d = 0.1~\mathrm{GeV}^2$. The threshold value for $\Re\{x\}$ is determined from 
		\begin{equation}
			x_{\mathrm{threshold}} = d / (1 - \zeta_0)
		\end{equation}
		in order for Eq.~\eqref{eq:root_finding_b} to yield a real solution. Meanwhile $\zeta_0$ in Eq.~\eqref{eq:root_finding_b} is the target center of the peak in the mapped variable $\zeta$, a typical choice of which is $0.5$. With this choice of $\zeta_0$ we have
		\begin{equation}
			b = 0.63678181 / ( 1.0 - z_0 )
		\end{equation} 
		as an approximate solution of Eq.~\eqref{eq:root_finding_b} for $z_0 \geq 0.9999$. While  for other values of $z_0$ the parameter $b$ can be solved from a root finder. Specifically we apply the method of bisection. Our choices of $d$ and $\zeta_0$ therefore gives $x_{\mathrm{threshold}} = 0.2\,\mathrm{GeV}^2$. The contribution to the self-energy from the peak is given by the $201$-point Gauss-Chebyshev quadrature of the second kind for the mapped variable $\zeta$, after subtracting the linear interpolations corresponding to the trapezoidal quadrature. The final result is a summation of results from both rules. 
		\item When $\Re\{x\} < x_{\mathrm{threshold}}$ the peak is wide and indistinguishable. In this case neither the trapezoidal rule nor the variable transformation is utilized. The $201$-point Gauss-Chebyshev quadrature of the second kind is instead applied to evaluate the integral for the original angular variable $z$. 
	\end{enumerate}
	After the angular integrals, we need to add the contribution to the self-energy from the asymptotics to obtain the inverse propagator as indicated by Eq.~\eqref{eq:sde_quark_prop_final_numeric}. The quark propagator is then given by its inverse based on Eq.~\eqref{eq:sigma_vs_from_AB}. 
	We have discussed the full algorithm in the computation of the quark propagator with $x\in\mathbf{C}$, which is illustrated by the activity diagram in Fig.~\ref{fig:fixedgridsf}. 
	\begin{figure}
		\centering
		\includegraphics[width=\linewidth]{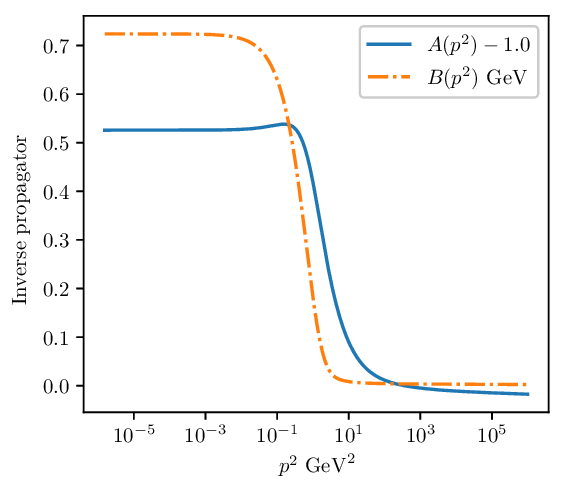}
		\caption{Inverse of the quark propagator as solution of the SDE in the Euclidean space from the iterative solver. Solid line: Dirac vector component after subtracting unity. Dash-dotted line: Dirac scalar component in units of $\mathrm{GeV}$.}
		\label{fig:sfeuc}
	\end{figure}
	\begin{table}
		\centering
		\begin{tabular}{ccccc}
			\hline
			$\Lambda^2_{\mathrm{UV}}$ & $\mu^2$ & $m_{\mathrm{R}}$ \\
			\hline
			$1.0\times 10^6\,\mathrm{GeV}^2$ & $361.0\,\mathrm{GeV}^2$ & $3.6964\,\mathrm{MeV}$ \\
			\hline
		\end{tabular}
		\caption{Parameters for the numerical solution of the SDE and the computation of the quark propagator with complex-valued momenta. The UV cutoff, the renormalization scale, and the renormalized mass are respectively given by $\Lambda^2_{\mathrm{UV}}$, $\mu^2$, and $m_{\mathrm{R}}$.}\label{tab:pmt_num}
	\end{table}
	\section{Accuracy and performance tests\label{sc:test}}
	With the model parameters in Table~\ref{tab:pmts_MT} for the Maris-Tandy model, we solve the SDE in the Euclidean space numerically with other physical parameters listed in Table~\ref{tab:pmt_num}. The variable transformation scale in Eq.~\eqref{eq:def_u_of_p2} is chosen to be ${C=10.0\,\mathrm{GeV}^2}$, after which the Gauss-Chebyshev quadrature of the first kind with ${N_{\mathrm{GC}}=2001}$ points is constructed for the Euclidean-space solution. We then compute the kernels of the iterative solver for the SDE in Eq.~\eqref{eq:sde_quark_prop_symmeetric_z}. The solution from the iterative solver with a tolerance of $1.0\times 10^{-6}$ is illustrated in Fig.~\ref{fig:sfeuc}. Both scalar functions are real valued for positive $p^2$. The obtained renormalization constants are ${Z_2 = 0.982\,009}$ and ${Z_{\mathrm{m}} = 0.670\,477}$. 
	\begin{table}
		\centering
		\begin{tabular}{cccc}
			\hline
			$N$ & Fixed grid & Adaptive grid \\
			\hline
			$36$ & $0.517$ s & $11.325$ s \\
			$72$ & $0.822$ s & $21.967$ s \\
			$144$ & $1.485$ s & $42.609$ s \\
			$288$ & $2.858$ s & $86.415$ s \\
			$576$ & $5.458$ s & $170.995$ s \\
			$1152$ & $10.941$ s & $339.946$ s \\
			\hline
		\end{tabular}
		\caption{Elapsed real time in the computation of the inverse quark propagator with complex-valued momenta for $N$ points in a geometric series of $k^2$. OpenMP is utilized to apply $36$ concurrent processes.}
		\label{tab:test}
	\end{table}
	
\begin{figure}
		\centering
		\includegraphics[width=\linewidth]{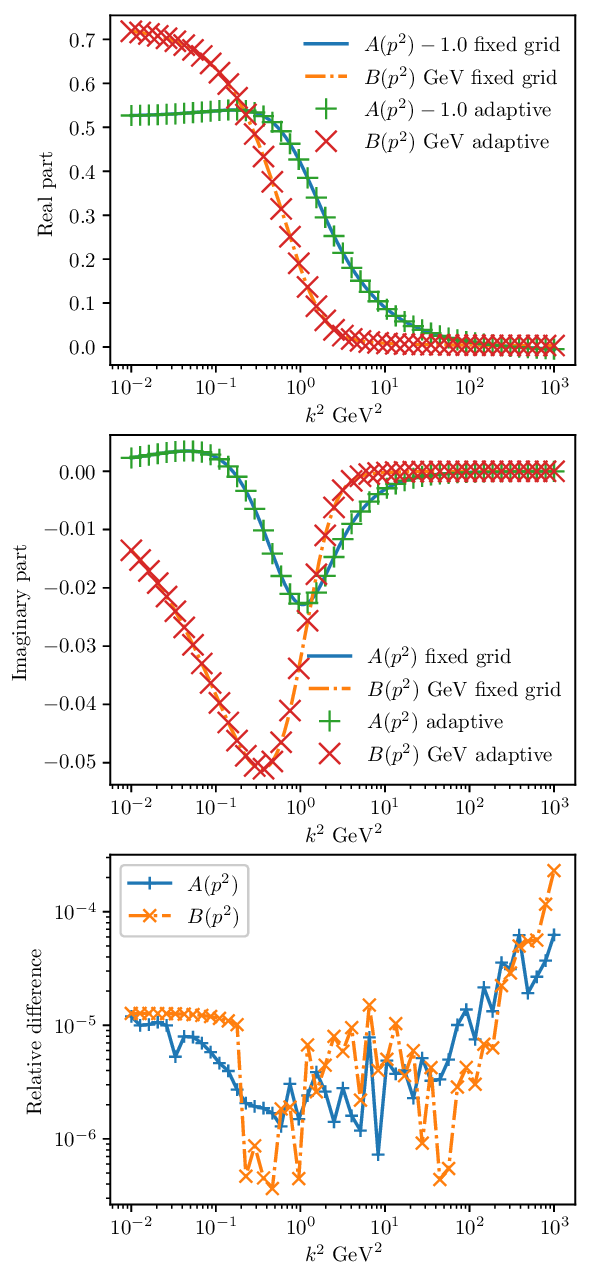}
		\caption{Accuracy test for the inverse of the quark propagator on the boundary of integrations sampled by the BSE for pions. Horizontal axes are the loop momenta. Top panel: real parts of the scalar components. Unity is subtracted from the Dirac vector components for illustration. Middle panel: imaginary parts of the scalar components. Blue solid lines and yellow dash-doted lines stand for results of $A(p^2)$ and $B(p^2)$ respectively from fixed-grid quadrature rules discussed in this article. Green plus signs and red crosses represent standard results from the adaptive Gauss-Kronrod quadrature for $A(p^2)$ and $B(p^2)$ respectively. Bottom panel: absolute values of the relative differences for results from the fixed-grid quadrature with respect to those from the adaptive quadrature. The blue solid line and the yellow dash-dotted line respectively stand for the absolute values of the relative differences in $A(p^2)$ and $B(p^2)$.}\label{fig:testsfcompiled}
	\end{figure}

	Based on the Euclidean-space solution, we then compute the quark propagator on the boundary of the BSE integration region corresponding to pions of mass $m_{\pi} = 137.24\,\mathrm{MeV}$. Specifically the sampled section in the complex-momentum plane is enclosed by following parabola:
	\begin{equation}
		\begin{cases}
			\Re\{p^2\} = k^2 - m^2_{\pi}/4 \\
			\Im\{p^2\} = \pm m_{\pi} k
		\end{cases},\label{eq:def_parabolic_bse}
	\end{equation}
	with $k^2$ being the square of the loop momentum and ${k = \sqrt{k^2}}$. We compute the inverse of the quark propagator at the boundary defined by Eq.~\eqref{eq:def_parabolic_bse} with a geometric series of $k^2$ applying the fixed-grid method described in this article. For the references of both accuracy and performance, the standard values of the inverse propagator are computed with the Gauss-Kronrod adaptive quadrature for both the radial and the angular integrals in Eq.~\eqref{eq:sde_quark_prop_symmeetric_z}. The real and imaginary parts of these results with the absolute values for their relative differences are illustrated in Fig.~\ref{fig:testsfcompiled}. With a tolerance chosen of $1.0\times 10^{-3}$ for the adaptive quadrature, the relative differences are below $7.2 \times 10^{-5}$ for the Dirac vector component and below $2.3 \times 10^{-4}$ for the Dirac scalar component. For a selection of $N$ points in the loop momentum $k^2$, the elapsed real time of these computation is given in Table~\ref{tab:test}.
	\section{Acknowledgments}
	This work was supported by the US Department of Energy, Office of Science, Office of Nuclear Physics, under Contract No. DE-AC02-06CH11357. We gratefully acknowledge the computing resources provided on Bebop, a high-performance computing cluster operated by the Laboratory Computing Resource Center at Argonne National Laboratory.
	\appendix
	\section{Gauss-Chebyshev quadrature\label{sc:quad_gc}}
	The Gauss-Chebyshev quadrature~\cite{abramowitz1965handbook} of the first kind is designed to compute the following integral with respect to $\xi$
	\begin{subequations}
		\begin{equation}
			\int_{-1}^{1}d\xi \, \dfrac{f(\xi)}{\sqrt{1-\xi^2}} \simeq w\sum_{j=1}^{N_{\mathrm{GC}}}f(\xi_j),
		\end{equation}
	with roots corresponding to the Chebyshev polynomial of the first kind such that
		\begin{equation}
			\begin{cases}
				w = \pi / N_{\mathrm{GC}} \\
				\xi_j = \cos( (j-1/2)\, w )
			\end{cases}.\label{eq:def_xi_quad_gc_I}
		\end{equation}
	\end{subequations}
	Here $N_{\mathrm{GC}}$ is the number of grid points. While the Gauss-Chebyshev quadrature rule of the second kind evaluates a different integral
	\begin{subequations}
		\begin{equation}
			\int_{-1}^{1}d\xi \, f(\xi)\sqrt{1-\xi^2} \simeq w\sum_{j=1}^{N_{\mathrm{GC}}}f(\xi_j),
		\end{equation}
	where roots are given by the Chebyshev polynomial of the second kind such that
		\begin{equation}
			\begin{cases}
				w = \pi / (N_{\mathrm{GC}}+1) \\
				\xi_j = \cos( jw )
			\end{cases}.\label{eq:def_xi_quad_gc_II}
		\end{equation}
	\end{subequations}
	
	These quadrature rules can be applied to compute the integral ${\int_{a}^{b}dx\,g(x)}$
	after a variable transformation
	\begin{equation}
		\xi = 2\, ( x - a ) / ( b - a ) - 1,
	\end{equation}
	which results in the inverse transformation of $x = {( b - a ) ( \xi + 1 ) / 2 + a}$. Explicitly we obtain
	\begin{equation}
		\int_{a}^{b}dx\,g(x) = \dfrac{b-a}{2}\int_{-1}^{1}d\xi\, g\left( ( b - a ) ( \xi + 1 )/2 + a \right).\label{eq:quad_gc}
	\end{equation}
	If we apply the quadrature rule of the first kind, the integral in Eq.~\eqref{eq:quad_gc} becomes
	\begin{align}
		\int_{a}^{b}dx\,g(x) & \simeq \dfrac{(b-a)\pi}{2N_{\mathrm{GC}}} \sum_{j=1}^{N_{\mathrm{GC}}}g\left( ( b - a ) ( \xi_j + 1 )/2 + a\right) \nonumber\\
		& \quad \times \sqrt{1-\xi^2_j} \label{eq:quad_gc_I}
	\end{align}
	with $\xi_j$ given by Eq.~\eqref{eq:def_xi_quad_gc_I}. When the quadrature rule of the second kind is applied, the integral becomes
	\begin{align}
		\int_{a}^{b}dx\,g(x) & \simeq \dfrac{(b-a)\pi}{2(N_{\mathrm{GC}}+1)} \sum_{j=1}^{N_{\mathrm{GC}}}g\left( ( b - a ) ( \xi_j + 1 )/2 + a\right) \nonumber\\
		& \quad \times \sqrt{1-\xi^2_j} \label{eq:quad_gc_II}
	\end{align}
	with $\xi_j$ given by Eq.~\eqref{eq:def_xi_quad_gc_II}. In both scenarios, the factor of $\sqrt{1-\xi^2}$ is contained in the function $g(x)$. An optimized choice of the quadrature type can be made based on the end-point behavior of $g(x)$. 
	
	The quadrature rules in Eqs.~\eqref{eq:quad_gc_I} and \eqref{eq:quad_gc_II} both indicate the following roots and weights for the variable ${x\in[a,b]}$
	\begin{subequations}\label{eq:roots_weights_mapped_quad_gc}
		\begin{align}
			x_{N_{\mathrm{GC}} - j + 1} & = (b - a) (\xi_j + 1) / 2 + a, \\
			w_{N_{\mathrm{GC}} - j + 1} & = \frac{b - a}{2} w \sqrt{1 - \xi^2_j},
		\end{align}
	\end{subequations}
	with $w$ and $\xi_j$ given by Eq.~\eqref{eq:def_xi_quad_gc_I} for the first kind and Eq.~\eqref{eq:def_xi_quad_gc_II} for the second kind. Equation~\eqref{eq:roots_weights_mapped_quad_gc} is applied to construct the grid for the mapped radial variable $u$ in Eq.~\eqref{eq:def_u_of_p2}. Indices in Eq.~\eqref{eq:roots_weights_mapped_quad_gc} are flipped to maintain monotonicity of the variable $x$ with respect to its indices. 
	\section{Adaptive Gauss-Kronrod quadrature\label{sc:quad_gk}}
	The following integral applies the Gauss-Kronrod rule
	\begin{equation}
		\int_{a}^{b}dx\,f(x) \simeq \dfrac{b-a}{2} \sum_{j=1}^{N_{\mathrm{QK}}} w_j \, f\left( (b - a) ( x_j + 1 ) / 2 + a \right), \label{eq:quad_gk}
	\end{equation}
	where $w_j$ and $x_j$ are respectively the roots and weights of the Gauss-Kronrod quadrature for the integral from $-1$ to $1$. The number $N_{\mathrm{GK}}$ stands for the number of grid points for the quadrature. We apply the adaptive quadrature rule where the integral in Eq.~\eqref{eq:quad_gk} is evaluated both by the Gauss rule and by the Kronrod rule. When their difference is below the desired tolerance, the result from the Kronrod rule corresponds the value of the integral in the working interval. Otherwise the interval is divided into ${[a, (a+b)/2]}$ and ${[(a+b)/2, b]}$ for further quadrature evaluations. The application of the adaptive quadrature is therefore achieved through function recursion. Specifically we apply the 7-point Gauss rule embedded in the 15-point Kronrod rule~\cite{quad_gk_rw}. 
	\section{Gauss-Hermite quadrature\label{sc:quad_gh}}
	The Gauss-Hermite quadrature gives the following approximation of the integral
	\begin{equation}
		\int_{-\infty}^{+\infty}du\,e^{-u^2}\,g(u) \simeq \sum_{j=1}^{N_{\mathrm{GH}}} \tilde{w}_j\, g(u_j),\label{eq:quad_gh_ori}
	\end{equation}
	with $N_{\mathrm{GH}}$ being the number of quadrature points, $u_j$ and $\tilde{w}_j$ the roots and the original weights~\cite{abramowitz1965handbook}. When the exponential is not factored from the integrand, we have ${f(u)} = {e^{-u^2} g(u)}$. In this scenario Eq.~\eqref{eq:quad_gh_ori} becomes
	\begin{equation}
		\int_{-\infty}^{+\infty}du\,f(u) \simeq \sum_{j=1}^{N_{\mathrm{GH}}} w_j \, f(u_j),\label{eq:quad_gh}
	\end{equation}
	where the weights are related to the original ones by $w_j = e^{u^2_j}\tilde{w}_j$. We apply the $51$-point Gauss-Hermite quadrature rule with roots $u_j$ and weights $w_j$ as follows. 
	\begin{verbatim}
		unsigned GH51::n_grid = 51;
		double GH51::r[] =
		{-9.284352965094795351319589826744050e+00,
			-8.627089729363683900942305626813322e+00,
			-8.082012673504012312264421780128032e+00,
			-7.594737944583593858283165900502354e+00,
			-7.144520571479617387922189664095640e+00,
			-6.720677095868753170293530274648219e+00,
			-6.316797657903404861201579478802159e+00,
			-5.928653885873137241446784173604101e+00,
			-5.553268005264917483998488023644313e+00,
			-5.188439342938137244232166267465800e+00,
			-4.832479614578074844644106633495539e+00,
			-4.484054079057350428172412648564205e+00,
			-4.142080801962062075460835330886766e+00,
			-3.805663856751933415978328412165865e+00,
			-3.474047361655858257023510304861702e+00,
			-3.146582843771695614520922390511259e+00,
			-2.822705423305429395242072132532485e+00,
			-2.501916004057736842014492140151560e+00,
			-2.183767652536192649392887688009068e+00,
			-1.867854955699560282056381765869446e+00,
			-1.553805529455421829965189317590557e+00,
			-1.241273096623611404965004112455063e+00,
			-9.299317156001758455374783807201311e-01,
			-6.194708497471526076338932398357429e-01,
			-3.095910409372403804262319226836553e-01,
			0.000000000000000000000000000000000e+00,
			3.095910409372403804262319226836553e-01,
			6.194708497471526076338932398357429e-01,
			9.299317156001758455374783807201311e-01,
			1.241273096623611404965004112455063e+00,
			1.553805529455421829965189317590557e+00,
			1.867854955699560282056381765869446e+00,
			2.183767652536192649392887688009068e+00,
			2.501916004057736842014492140151560e+00,
			2.822705423305429395242072132532485e+00,
			3.146582843771695614520922390511259e+00,
			3.474047361655858257023510304861702e+00,
			3.805663856751933415978328412165865e+00,
			4.142080801962062075460835330886766e+00,
			4.484054079057350428172412648564205e+00,
			4.832479614578074844644106633495539e+00,
			5.188439342938137244232166267465800e+00,
			5.553268005264917483998488023644313e+00,
			5.928653885873137241446784173604101e+00,
			6.316797657903404861201579478802159e+00,
			6.720677095868753170293530274648219e+00,
			7.144520571479617387922189664095640e+00,
			7.594737944583593858283165900502354e+00,
			8.082012673504012312264421780128032e+00,
			8.627089729363683900942305626813322e+00,
			9.284352965094795351319589826744050e+00};
		double GH51::w[] =
		{7.586846833851778049151448612974491e-01,
			5.864287425038513523745109523588326e-01,
			5.113148070267715672443387120438274e-01,
			4.664299739336288941338182212348329e-01,
			4.357071372202365178161187486693962e-01,
			4.130175315947753333922776164399693e-01,
			3.954318931792801250502122911711922e-01,
			3.813445580868277118291587157727918e-01,
			3.697897868419675604378937805449823e-01,
			3.601463743888477786470048158662394e-01,
			3.519939135068397839667397875018651e-01,
			3.450363958905212635741577287262771e-01,
			3.390587302852089979232630412298022e-01,
			3.339006067526961718527900302433409e-01,
			3.294400717697356650859319415758364e-01,
			3.255828179885981166208352988178376e-01,
			3.222549814252920197255036782735260e-01,
			3.193981705344928201384391286410391e-01,
			3.169659611758387507407519478874747e-01,
			3.149213821221034659281201584235532e-01,
			3.132350877650721399092503816063982e-01,
			3.118840198763703774886835162760690e-01,
			3.108504266084930467428648626082577e-01,
			3.101211499903285817580922412162181e-01,
			3.096871220156214854490883681137348e-01,
			3.095430294466409715248289558076067e-01,
			3.096871220156214854490883681137348e-01,
			3.101211499903285817580922412162181e-01,
			3.108504266084930467428648626082577e-01,
			3.118840198763703774886835162760690e-01,
			3.132350877650721399092503816063982e-01,
			3.149213821221034659281201584235532e-01,
			3.169659611758387507407519478874747e-01,
			3.193981705344928201384391286410391e-01,
			3.222549814252920197255036782735260e-01,
			3.255828179885981166208352988178376e-01,
			3.294400717697356650859319415758364e-01,
			3.339006067526961718527900302433409e-01,
			3.390587302852089979232630412298022e-01,
			3.450363958905212635741577287262771e-01,
			3.519939135068397839667397875018651e-01,
			3.601463743888477786470048158662394e-01,
			3.697897868419675604378937805449823e-01,
			3.813445580868277118291587157727918e-01,
			3.954318931792801250502122911711922e-01,
			4.130175315947753333922776164399693e-01,
			4.357071372202365178161187486693962e-01,
			4.664299739336288941338182212348329e-01,
			5.113148070267715672443387120438274e-01,
			5.864287425038513523745109523588326e-01,
			7.586846833851778049151448612974491e-01};
	\end{verbatim}
	\bibliographystyle{elsarticle-num}
	\bibliography{quark_self_energy_bib}
	%
	%
		%
		%
		%
\end{document}